\documentclass[10pt,prd,twocolumn]{revtex4-1}
\usepackage{hyperref,graphicx,color,amsmath, amssymb,natbib, url}

\def\jcap{J. Cosmol. Astropart. Phys.}%
\def\prd{Phys.~Rev.~D}%
\def\aap{A\&A}%

\def\vecTheta {\mathbf{\Theta}}
\def\vecO {\mathbf{O}}

\begin{document}
\title{Marginalized Fisher Forecast for Horndeski Dark Energy Models}
\author{Jason S.-Y. Leung$^{1}$, Zhiqi Huang$^{2}$}
\affiliation{
${}^1$Department of Astronomy and Astrophysics, University of Toronto, 50 St.~George Street, Toronto, Ontario, M5S 3H4, Canada \\
${}^2$Canadian Institute for Theoretical Astrophysics, University of Toronto, 60 St.~George Street, Toronto, Ontario, M5S 3H8, Canada}

\date{\today}

\begin{abstract}
  
We use effective field theory (EFT) formalism to forecast the constraint on Horndeski class of dark energy models with future supernova and galaxy surveys. Previously (Gleyzes {\it et al.}) computed unmarginalized constraints (68\% CL error $\sim 10^{-3}$--$10^{-2}$) on EFT dark energy parameters by fixing all other parameters. We extend the previous work by allowing all cosmological parameters and nuisance parameters to vary and marginalizing over them. We find that (i) the constraints on EFT dark energy parameters are typically worsen by a factor of few after marginalization, and (ii) the constraint on the dark energy equation of state $w$ is not significantly affected by the inclusion of EFT dark energy parameters.
  
\end{abstract}

\maketitle

\section{Introduction}

Following the discovery of cosmic acceleration two decades ago, a flurry of theoretical work has culminated in what is now known as the $\Lambda$CDM model, or the concordance model. By introducing a cosmological constant term ($\Lambda$) in addition to the presence of cold dark matter (CDM), it is able to broadly account for a large number of cosmological observations: galaxy power spectra, weak gravitational lensing, cluster counting, Lyman-alpha forest, type Ia supernova light-curves, and the cosmic microwave background (CMB) anisotropy. The refinement of its six parameters is the subject of numerous present and future scientific experiments.

The most recent data from the \textit{Planck} satellite shows broad consistency with the $\Lambda$CDM predictions \cite{Planck2015}. However, detailed analysis of the observed CMB temperature and polarization anisotropies reveals hints of deviations from these predictions: a temperature power spectrum deficit at large angular scales (multipole $\ell \lesssim 30$) and a lensing potential power spectrum excess at small angular scales~\cite{PlanckIandS}. Furthermore, local measurements suggest a higher Hubble constant $H_0$ and a lower amplitude of r.m.s. matter density fluctuations $\sigma_8$ compared to those suggested by \textit{Planck}~\cite{Riess/etal:2011,Verde/etal:2013,Planck2015,Bohringer/etal:2015}.

While the anomalies each only lie at the $\sim\!2\sigma$ level and hence are not of great statistical significance, \cite{Bellini2016} found them ``persistent.'' One possible explanation for this is that general relativity breaks down at cosmological scales, requiring modifications to the theory of gravity to account for the deviations observed. This hypothesis is attractive not only because it circumvents the need for post-hoc constructs such as the cosmological constant, but also because the discovery of a single model accounting for all of the anomalies would open up the possibility of ``new physics'' beyond the standard $\Lambda$CDM model. Much ink has already been spent in pursuit of such ``modified gravity'' models, but the continuing absence of strong theoretical priors has left the search for a ``real'' theory of gravity largely fruitless.

In order to make sense of the plethora of models existing in the literature, it would be worthwhile to devise a common language to express them. This would facilitate the comparison of different models against one another, as well as provide a generic framework around which new models could be built. To this end, effective field theory (EFT) has emerged as a prime contender in providing a unifying description of dark energy \cite{Creminelli/etal:2009,Gubitosi2013, Bellini2014, Bloomfield2013, Gleyzes/etal:2013, Gleyzes2014, Gleyzes/etal:2015}.

Linear perturbations upon a spatially flat Friedmann--Lema\^{i}tre--Robertson--Walker (FLRW) background can be described in the EFT language using five functions dependent only on time: the dark energy equation of state $w$, the effective Planck mass run rate $\alpha_M$, the tensor speed excess $\alpha_T$, the kineticity $\alpha_K$, and the scalar--tensor braiding $\alpha_B$~\cite{Bellini2014, Gleyzes2014}. In theories beyond Horndeski gravity, a sixth function representing these deviations, $\alpha_H$, also appears  \cite{Gleyzes/Langlois/etal:2015, GLPV2015}. The full forms of the $\alpha_X$ (X = B, M, T, K) functions can be found in Table 1 of Ref.~\cite{Gleyzes2014}, while their physical interpretations are described in \S2.1 of Ref.~\cite{Bellini2016}. In the $\Lambda$CDM model, $w \equiv -1$ while $\alpha_X \equiv 0$.

Constraints on these parameters based on current cosmological data were calculated by Ref.~\cite{Bellini2016}, while Fisher matrices were calculated by Ref.~\cite{Gleyzes2016}. Due to the highly degenerate nature of the EFT parameters ($w$, $\alpha_X$) with other cosmological parameters, Ref.~\cite{Gleyzes2016} completed their Fisher matrix analysis by fixing all cosmological and nuisance parameters to their fiducial values. Without marginalization, 68\% confidence level (CL) errors were found to be on the order of $\sigma \sim 10^{-3}$--$10^{-2}$. In this paper, we extend their work by allowing all cosmological and nuisance parameters to vary and then marginalizing over them.

\section{Method}

For a set of parameters
\begin{equation}
\vecTheta = \left(\theta_1, \theta_2, \ldots, \theta_m\right) \,
\end{equation}
and a set of observables
\begin{equation}
  \vecO \!\left(\vecTheta\right)= \left(O_1\!\left(\vecTheta\right), O_2\!\left(\vecTheta\right), \ldots, O_n\!\left(\vecTheta\right)\right) \,,
\end{equation}
knowing the Fisher matrix for $\vecO$ allows one to write the Fisher matrix for $\vecTheta$ as
\begin{equation}
F_{\theta_i,\theta_j} = \sum_{k,l}\frac{\partial O_k}{\partial \theta_i} F_{O_k,O_l} \frac{\partial O_l}{\partial \theta_j} \,.
\end{equation}

In cosmology, the observables $O$ are the CMB multiples $C_l^{TT}, C_l^{TE}, C_l^{EE}$, the galaxy power spectrum in redshift space~\cite{Kaiser:1987, Peacock:1992, Peacock/Dodds:1994}, and the magnitude of type Ia supernovae.

The foreground-cleaned CMB power spectrum is modeled as
\begin{equation}
  \hat{C}_\ell = b_\ell^2 C_\ell^\mathrm{CMB} + A_\mathrm{noise} C_l^\mathrm{noise}  \, ,
\end{equation}
where the beam window function is assumed to be Gaussian, such that $b_\ell = \exp\!\left[- \theta_b^2\ell\left(\ell+1\right)\!/2\right]$. The beam resolution is often specified by its full width at half maximum (FWHM), which equals to $\sqrt{8\ln 2}\,\,\theta_b$. The noise is assumed to be Gaussian and white. The overall scaling parameter $A_\mathrm{noise}$ accounts for the uncertainty of noise power estimation. We assume a $1\%$ level accuracy of noise estimation and use a Gaussian prior $A_\mathrm{noise} = 1 \pm 0.01$.

For the CMB Fisher matrix, we use \textit{Planck}'s 70~GHz, 100~GHz and 143~GHz channels. The specification of noise level and beam size of each channel are specified in, for instance, Table I of Ref.~\cite{Huang2012}. But while \textit{Planck} data has been partially released and a likelihood is publicly available~\cite{PlanckLikelihood}, the current \textit{Planck} low-$\ell$ polarization data is still being analyzed and the final constraint on cosmology is not yet known. Thus, we use a simplified Fisher matrix analysis here as an ideal case simulating a future (improved) constraint of the \textit{Planck} CMB.

The observed galaxy power spectrum is modeled as
\begin{equation}
P_g\!\left(k,\mu; z\right) = \left(b+ f \mu^2\right)^2 D^2\!\left(z\right)P_\mathrm{m}\!\left(k\right) e^{-k^2\mu^2\sigma_r^2} + \frac{1}{\bar{n}}, \label{eq:Pg}
\end{equation}
where $z$ is the redshift, $\mu$ is the cosine of the angle between the wavenumber $\mathbf{k}$ and the line of sight, $b$ is the galaxy bias, $D(z)$ is the linear growth factor, $f \equiv -{\mathrm{d} \ln D}/{\mathrm{d} \ln \left(1+z\right)}$ is the linear growth rate, $P_\mathrm{m}\!\left(k\right)$ is the matter power spectrum today (at redshift $z=0$) and $\sigma_r$ is the parameterization of the effect of small-scale velocity dispersion and redshift errors:
\begin{equation}
\sigma_r^2 = \frac{\left(1+z\right)^2}{H^2\!\left(z\right)}  \left(\sigma_z^2 + {\sigma_g^2}/{2} \right)\;, \label{eq:sigma2}
\end{equation}
where  $H(z)$ is the Hubble parameter as a function of redshift. Following Ref.~\cite{Huang2012}, we consider a future galaxy survey with spectroscopic $\sigma_z = 0.001$. The galaxy pairwise velocity dispersion $\sigma_g$ is treated as a nuisance parameter with Gaussian prior $\sigma_g = 400 \pm 200 \,\,\mathrm{km\,s^{-1}Mpc^{-1}}$. We use 8 redshift bins from $z=0.5$ to $z=2.1$ with uniform bin size $\delta z = 0.2$. In each redshift bin we apply an IR cut $k>k_{\min}$ to account for the band limit due to a finite survey volume, and a UV cut $k<k_{\max}$ to guarantee that only linear scales are used. Bias in each redshift bin is considered as a nuisance parameter with flat prior $0<b<\infty$. The mean galaxy number density $\bar{n}$ which gives the poisson noise term $1/\bar{n}$ is assumed to be known. For all redshift bins, we assume a sky coverage $f_\mathrm{sky} = 0.5$. More details about the specifications are provided in Ref.~\cite{Huang2012}. 

The type Ia supernova apparent magnitude is 
\begin{equation}
M_\mathrm{obs} = 25 + M_\mathrm{abs} + 5 \log_{10}\!\left(d_L/\mathrm{Mpc}\right)\, ,
\end{equation}
where $M_\mathrm{abs}$ is the absolute magnitude for which we assume a Gaussian prior $M_\mathrm{abs} = 19 \pm 0.09$ mags.

We assume a \textit{WFIRST}-like observation with 2725 supernovae up to redshift $1.7$~\cite{Spergel2015}. The overall statistical error in a $\Delta z = 0.1$ redshift bin is modeled as:
\begin{equation}
\sigma_\mathrm{stat}^2 = \sigma^2_\mathrm{measure} + \sigma^2_\mathrm{intrinsic} + \sigma_\mathrm{lensing}^2 + \sigma_\mathrm{pec}^2 \, ,
\end{equation}
where the photometric measurement error $\sigma_\mathrm{measure} = 0.08$ mags, the intrinsic dispersion in type Ia supernova absolute magnitude $\sigma_\mathrm{intrinsic} = 0.09$ mags, the statistical error due to gravitational lensing magnification $\sigma_\mathrm{lensing} = 0.07 z$ mags, and the uncertainty due to supernova peculiar velocity $\sigma_\mathrm{pec} = 5 \log_{10} \!\left[v/\!\left(cz\right)\right]$ with a typical velocity $v = 400\,\,\mathrm{km\,s^{-1}}$. Finally, we add a systematic uncertainty $\sigma_\mathrm{sys} = 0.0055\left(1+z\right)$ to each redshift bin to account for calibration errors between different bins~\cite{Spergel2015}.

Within modified gravity models, there are several ways to define the ``equation of state'' $w$ of dark energy. We use a definition based on the background expansion history in the Jordan frame:
\begin{equation}\begin{split}
H^2\!\left(z\right) &= H_0^2 \left[ \Omega_{m0} \left(1+z\right)^3 + \Omega_{r0} \left(1+z\right)^4 \right. \\
& \left. + (1-\Omega_{m0}-\Omega_{r0}) e^{-3\int \left(1+w\right)\mathrm{d}z/\left(1+z\right)} \right]  \, , \\
\end{split}
\end{equation}
where $\Omega_{m0} \,\, \left( \Omega_{r0} \right)$ is the ratio between matter (relativistic) energy density and the critical density $\rho_{\rm crit} \equiv 3H_0^2/(8\pi G_N)$, $G_N$ being Newton's gravitational constant and $H_0$ the Hubble constant (i.e., the Hubble parameter today). In general, a subscript ``0'' denotes a quantity measured at redshift zero.

We take three benchmark examples to show the effect of marginalization: (A) fixing all other cosmological parameters and nuisance parameters; (B) fixing other cosmological parameters and only marginalizing over nuisance parameters; (C) marginalizing over all other cosmological parameters and nuisance parameters.  Model A corresponds to an unmarginalized constraint and can be directly compared to Ref.~\cite{Gleyzes2016}.  Model B is an ideal case assuming that additional experiments (e.g., CMB stage IV \cite{CMBStageIV}) will measure other cosmological parameters to a much better accuracy.  Model C is a fully marginalized forecast using the aforementioned datasets.

We follow Ref.~\cite{Gleyzes2016} in parameterizing EFT dark energy: The equation of state $w$ is assumed to be a constant, while the other EFT functions are parameterized as
\begin{equation}\begin{split}
\alpha_K \!\left(z\right) &= \alpha_{K0} \frac{\rho_\mathrm{DE}\!\left(z\right)}{\rho_\mathrm{DE,0}} \, , \\
\alpha_M \!\left(z\right) &= \alpha_{M0} \frac{\rho_\mathrm{DE}\!\left(z\right)}{\rho_\mathrm{DE,0}} \, , \\
\alpha_B \!\left(z\right) &= \alpha_{B0} \frac{\rho_\mathrm{DE}\!\left(z\right)}{\rho_\mathrm{DE,0}} \, , \\
\alpha_T \!\left(z\right) &= \alpha_{T0} \frac{\rho_\mathrm{DE}\!\left(z\right)}{\rho_\mathrm{DE,0}} \, . \\
\end{split}\end{equation}
Finally, we take $\alpha_H \!\left(z\right) \equiv 0$ in accordance with Horndeski gravity models. For simplicity, the dark energy fraction density ratio $\rho_\mathrm{DE}(z) / \rho_\mathrm{DE,0}$ is defined for a fixed $\Lambda$CDM model with $\Omega_{m0} = 0.3$.  Note that each $\alpha_{X0}$ may be fixed to fiducial values or allowed to vary. We therefore consider two cases: (I), where we vary each $\alpha_{X0}$ and $w$ individually and keep the remaining fixed, and (II), where we vary all $\alpha_{X0}$ and $w$ simultaneously. The standard deviation $\sigma$, representing the 68\% CL error, is then used as a measure of the constraint upon each parameter (or equivalently, its uncertainty). Standard deviations in case II are determined by marginalizing over all other $\alpha_{X0}$ and $w$. Consequently, any dependencies or correlations amongst the EFT parameters can be revealed by simply comparing the constraints upon them between the two cases. 

We use the Cosmology Object Oriented Package (COOP)~\cite{Huang2016} to compute the Fisher matrices. The list of parameters is given in Table~\ref{tbl:params}.

\begin{table*}
\caption{Fisher Forecast Parameters \label{tbl:params}}
\begin{tabular}{lllll}
\hline
\hline
Parameter & Genre & Physical interpretation & Fiducial & Prior \\ \hline
$\Omega_b h^2$ & cosmological & Baryon physical density & $0.02222$ & flat \\
$\Omega_c h^2$ & cosmological & CDM physical density & $0.1199$  & flat \\
$\theta$ & cosmological & Angular extension of sound horizon at recombination & $0.010486$ & flat \\
$\tau$ & cosmological & Reionization optical depth & $0.078$ & flat \\
$\ln (10^{10}A_s)$ & cosmological &  Logarithmic amplitude of primordial power spectrum& $3.09$ & flat \\
$n_s$ & cosmological & Spectral index of primordial power spectrum & $0.9655$ & flat \\
$w$ & cosmological & Dark energy equation of state & $-1$ & flat \\
$\alpha_{M0}$ & cosmological & Effective Planck mass run rate & $0$ & flat \\
$\alpha_{B0}$ & cosmological & Tensor-scalar braiding & $0$ & flat \\
$\alpha_{K0}$ & cosmological & Kineticity & $0.1$ & flat \\
$\alpha_{T0}$ & cosmological & Tensor speed excess & $0$ & flat \\
$M_*$ & nuisance & Type Ia supernova absolute magnitude & $19$ & $\pm 0.09$ \\
$A_{\rm noise}$ & nuisance & Amplitude of CMB noise power spectrum & $1$ & $\pm 0.01$ \\
$\sigma_g$  & nuisance & Galaxy pairwise velocity dispersion & $400$ km\,s$^{-1}$ & $\pm 200$ km\,s$^{-1}$ \\
$b_1$  & nuisance & Galaxy bias in redshift bin $0.5<z<0.7$ & $1.053$ & flat \\
$b_2$  & nuisance & Galaxy bias in redshift bin $0.7<z<0.9$ & $1.125$ & flat \\
$b_3$  & nuisance & Galaxy bias in redshift bin $0.9<z<1.1$ & $1.126$ & flat \\
$b_4$  & nuisance & Galaxy bias in redshift bin $1.1<z<1.3$ & $1.243$ & flat \\
$b_5$  & nuisance & Galaxy bias in redshift bin $1.3<z<1.5$ & $1.292$ & flat \\
$b_6$  & nuisance & Galaxy bias in redshift bin $1.5<z<1.7$ & $1.497$ & flat \\
$b_7$  & nuisance & Galaxy bias in redshift bin $1.7<z<1.9$ & $1.491$ & flat \\
$b_8$  & nuisance & Galaxy bias in redshift bin $1.9<z<2.1$ & $1.568$ & flat \\
\hline
\end{tabular}
\end{table*}

\section{Results}

For a dynamic model, the kineticity $\alpha_K$ usually does not vanish. We find that (i) the constraint on $\alpha_{K0}$ is very weak (uncertainty $> \mathcal{O}\!\left(1\right)$), and (ii) the forecast of other parameters is not sensitive to the fiducial value of $\alpha_{K0}$ (for $\alpha_{K0} \ll 1$). For small $\alpha_{B0}$, the constraint on $\alpha_{T0}$ is also very weak (see also \S6.2.1 of Ref.~\cite{Gleyzes2016}). Thus, we fix these two parameters to their fiducial values $\alpha_{K0} = 0.1$ and $\alpha_{T0} = 0$, and only discuss the constraints on $\alpha_{M0}$, $\alpha_{B0}$, and $w$. The standard deviations upon these three EFT parameters are summarized in Table \ref{tbl:models}.

In all cases, we observe a worsening of the constraints (i.e., $\sigma$ increases) as we progress from model A through C by marginalizing over an increasing number of parameters. Notice that the standard deviations increase by half an order of magnitude for every parameter in case I and for $w$ in case II. By contrast, the increase in uncertainty for $\alpha_{M0}$ and $\alpha_{B0}$ in case II appears much less pronounced.

This is not unexpected, for $w$ is in principle degenerate with all background and perturbative parameters, whereas the $\alpha_X$ are only degenerate with the latter. As a result, much of the uncertainty associated with $w$ is contained in the cosmological and nuisance parameters. Having already marginalized over these parameters in case I of model C, we therefore expect that marginalizing over the remaining EFT parameters ($w$ in particular) in case II should not increase the uncertainties on $\alpha_{X0}$ much further -- as if they have reached a ``saturation'' point.

\begin{figure}
\includegraphics[width=0.95\linewidth]{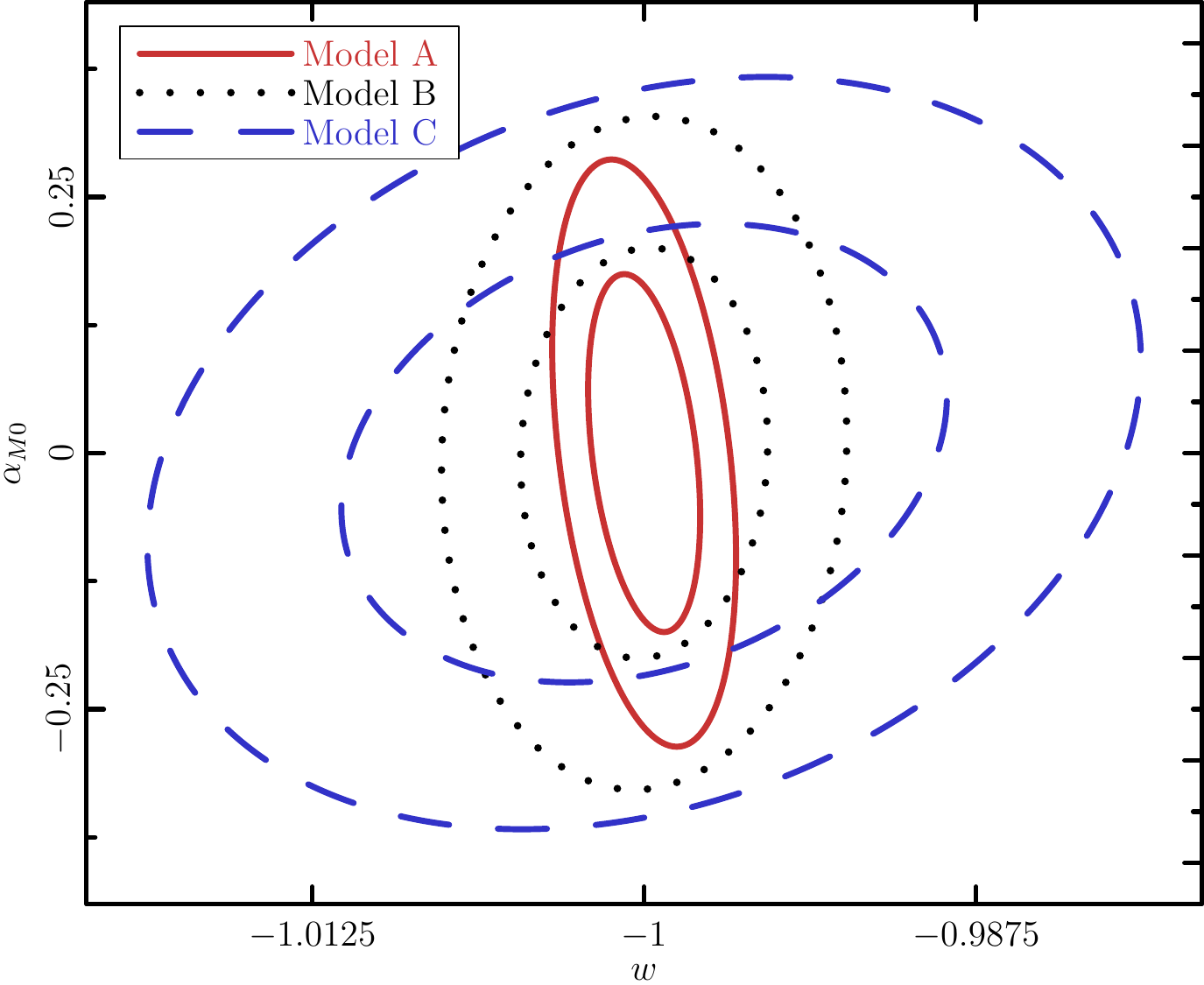}
\caption{Constraints on the effective Planck mass run rate $\alpha_{M0}$ and dark energy equation of state $w$ in case II, calculated with all parameters unmarginalized (Model A), nuisance parameters marginalized (Model B), and all parameters fully marginalized (Model C). Each inner ellipse corresponds to the 68\% confidence level error and the outer to the 95\%.}
\label{fig:contours}
\end{figure}

Finally, introducing additional degrees of freedom can change the correlations between parameters. An example can be seen in the contour plot of $\alpha_{M0}$ against $w$ (Figure \ref{fig:contours}): In addition to projecting much larger error ellipses, the sign of the correlation also flips from negative to positive after allowing all parameters to vary (model C).

\begin{table}
\caption{Constraints on $\alpha_{M0}$, $\alpha_{B0}$, and $w$ for two cases of the three models considered. Shown are the standard deviations, representing the 68\% CL errors on each parameter.}
\label{tbl:models}
\begin{tabular}{lllll}
\hline
\hline
Case & Parameter & Model A & Model B & Model C \\
\hline
      & $\alpha_{M0}$ &  $0.022$ & $0.080$ & $0.101$ \\
I     & $\alpha_{B0}$ &  $0.016$ & $0.049$ & $0.059$ \\
      & $w$ & $0.0008$  & $0.0031$ & $0.0069$ \\
\hline
       &  $\alpha_{M0}$  & $0.12$ & $0.13$ & $0.15$ \\
II     &  $\alpha_{B0}$  & $0.081$ & $0.082$ & $0.084$ \\
       &  $w$          & $0.0014$ & $0.0031$ & $0.0076 $ \\
\hline
\hline
\end{tabular}
\end{table}

\section{Conclusions}
By considering three cases with different levels of marginalization, we have shown that marginalization can weaken the constraints on the EFT parameters by a factor of a few. Results for the first case of model A can be directly compared with Ref.~\cite{Gleyzes2016}, which also projected $\sigma \sim 10^{-3}$--$10^{-2}$ for unmarginalized 68\% CL errors using a different dataset. This value worsened by a few factors after we marginalized over the nuisance parameters (model B), and worsened more so after marginalizing over all cosmological and nuisance parameters (model C).

Our constraint on the dark energy equation of state $w$ is on par with the level of uncertainty achievable with the most advanced upcoming probes ($\sigma_w \sim 10^{-2}$ in a $w$CDM model). Critically, allowing the $\alpha_{X0}$ to vary weakens the constraint on $w$ by less than $10^{-3}$ -- well within experimental error -- meaning that $w$CDM forecasts for future experiments remain valid. Since these upcoming probes already edge on the cosmic variance limit for $w$, we must also constrain $\alpha_{X0}$ to distinguish between a $w$CDM and a modified gravity universe. To that end, we obtain uncertainties of $\sigma_{\alpha_{M0}} \sim 10^{-1}$ and $\sigma_{\alpha_{B0}} \sim 10^{-2} $. For comparison, \textit{Planck} currently achieves an uncertainty on the order $\sigma \sim 1$ \cite{PlanckDE}.

That the forthcoming generation of dark energy probes is approaching the cosmic variance limit is important for another reason: As we approach that limit, cosmic variance itself becomes a significant part of the uncertainties. The standard approach for reducing these uncertainties and breaking degeneracies is to combine different cosmological observations such as galaxy and weak lensing power spectra and the ISW--galaxy-distribution correlation. Since these experiments all observe the same sky, the cosmic variance overlaps and so their Fisher matrices are correlated. In other words, these Fisher matrices cannot be added trivially, and proper treatment requires significant analysis of the relevant datasets to remove the correlations. We leave this analysis to future work.


\end{document}